\documentclass[twoside,twocolumn]{article}

\usepackage{blindtext} 

\usepackage[sc]{mathpazo} 
\usepackage[T1]{fontenc} 
\usepackage{microtype} 

\usepackage[english]{babel} 

\usepackage[hmarginratio=1:1,top=32mm,columnsep=20pt]{geometry} 
\usepackage[hang, small,labelfont=bf,up,textfont=it,up]{caption} 
\usepackage{booktabs} 

\usepackage{lettrine} 

\usepackage{enumitem} 
\setlist[itemize]{noitemsep} 

\usepackage{abstract} 

\usepackage{titlesec} 
\renewcommand\thesection{\Roman{section}} 
\renewcommand\thesubsection{\roman{subsection}} 
\titleformat{\section}[block]{\large\scshape\centering}{\thesection.}{1em}{} 
\titleformat{\subsection}[block]{\large}{\thesubsection.}{1em}{} 

\usepackage{fancyhdr} 
\usepackage{titling} 

\usepackage{hyperref} 
\usepackage{amsmath,times,graphicx,geometry,hyperref,amsthm , amssymb}
\usepackage{subfigure} 
 \usepackage{epsfig}
 \usepackage[section]{placeins}
 \usepackage{afterpage}

\pretitle{\begin{center}\Huge\bfseries} 
\posttitle{\end{center}} 
\title{Geometry of Quantum Riemannian Hamiltonian
	Evolution} 
\author{%
\textsc{Gil Elgressy,$^1$ Lawrence Horwitz,$^{1,2,3}$ }\\
\normalsize $^1$Department of Physics, Bar Ilan University, Ramat Gan 52900, Israel\\  
\normalsize $^2$Department of Physics, Ariel University, Ariel 44837, Israel\\
\normalsize $^3$School of Physics, Tel Aviv University, Ramat Aviv 69978, Israel\\
}
\date{\today} 
\renewcommand{\maketitlehookd}{%
	This work concerns a study of the quantum mechanical extension of  the work of Horwitz et al. \cite{horwitz} on the stability of classical Hamiltonian systems by geometrical methods.  
	Simulations are carried out for several important examples; these show that the quantum mechanical extension of the classical method, for which trajectories are plotted as expectation values of the corresponding quantum operators, appears to  work well, providing results consistent with the corresponding classical problems. The results appear to provide a new contribution to the subject of quantum chaos. 
}

\begin{document}

\maketitle







\section{Introduction}
A classical Hamiltonian of the standard form
\begin{equation}
H=\delta_{ij}\frac{p^ip^j}{2m}+V(\textbf{y})
\end{equation}
can be cast into the Riemannian form \cite{horwitz}
\begin{equation}
H_G=\frac{1}{2m}g_{ij}(\textbf{x})p^ip^j
\end{equation}
by choosing
\begin{equation}
g_{ij}(\textbf{x})=\Phi(\textbf{x})\delta_{ij}
\end{equation}
with, for $H$ and $H_G$ on the same energy shell E, the constraint
\begin{equation}
\Phi(\textbf{x})(E-V(\textbf{y}))={E}
\end{equation}
It follows from the classical Hamilton equations that \cite{gutzwiler}\cite{miler} the motion
generated by the Riemannian Hamiltonian (2) satisfies the geodesic relation
\begin{equation}
\ddot{x}_l=-\Gamma^{mn}_l\dot{x}_m\dot{x}_n
\end{equation}
where the connection form is given by (here $g^{ij}$ 
is the inverse of $g_{ij}$)
\begin{equation}
\Gamma^{mn}_l=\frac{1}{2}g_{lk}\{\frac{\partial g^{km}}{\partial x_n}+\frac{\partial g^{kn}}{\partial x_m}-\frac{\partial g^{nm}}{\partial x_k}\}
\end{equation}
Let us now define the variable $y^j$ for which
\begin{equation}
\dot{y}^j\equiv g^{ji}(\textbf{x})\dot{x}_i=\frac{p^j}{m}
\end{equation}
The last equality follows from the Hamilton equations based on the Riemannian Hamiltonian (2), and  the geodesic equation (5) implies that the variables \{\textbf{y}\} then satisfy the equation
\begin{equation}
\ddot{y}^l=-M^l_{mn}\dot{y}^m\dot{y}^n
\end{equation}
where
\begin{equation}
M^l_{mn}\equiv\frac{1}{2}g^{lk}\frac{\partial g_{nm}}{\partial y^k}
\end{equation}
corresponds to an affine connection governing the motion in the \{\textbf{y}\} description. 
Application of the constraint (4) to the definition (9) results in precisely the Hamilton equations derived from (1). Therefore, the evolution (8) corresponds to a geometric embedding of the usual Hamiltonian motion described by (1). 
With (7) a functional correspondence can be established between the variables $x_i$ and $y^j$ \cite{yahalom}.\\
An equivalence principle becomes accessible for the Hamiltonian (1) through the mapping defined by (3) and (4), since the connection form (5) is compatible with the metric (3).
Moreover, it is shown in \cite{horwitz} that the "geodesic deviation" computed from
(8) results in a Jacobi equation for which the eigenvalues associated with the resulting curvature tensor, constructed from the M connection (9), have remarkable diagnostic value in identifying stability properties for the system described by (1).\\
 In this work 
 we study the  quantum
theory associated with a general operator valued Hermitian Riemannian 
Hamiltonian 
\begin{equation}
\hat{H}_G=\frac{1}{2m}p^ig_{ij}(\textbf{x})p^j
\end{equation}
We shall show here that the variables corresponding to \{{\textbf{x}}\} in the Heisenberg picture, satisfy dynamical
equations closely related to those to the classical system, and therefore that when the Ehrenfest correspondence is valid, the expectation values of these variables would describe a corresponding flow.\\
We then construct the quantum counterpart of the relations (7), (8), (9) and show that these results are related to the quantum dynamics associated with a Hamilton operator of the form (1).\\
First we study the basic operator properties of coordinate and momentum
observables associated with a Hamiltonian operator of type (10) (with $g_{ij}=g_{ji}$ invertible).
Thus the canonical momentum is defined, both classically and quantum mechanically, as the generator of translation on a Euclidean space, and thus has a Schrodinger (coordinate) representation as a simple derivative.
\section{Heisenberg Algebra on The Geometric Hamiltonian}
The coordinates $\{x_i\}$ form a commuting set, as do the $\{p^j\}$.
The Heisenberg equations for the coordinates ($\hbar=1$);
\begin{equation}
\dot{x}_k=i[\hat{H}_G, x_k]
\end{equation}
with the canonical commutation relations
\begin{equation}
[x_i, p^j]=i\delta_i^j
\end{equation}
result in
\begin{equation}
\dot{x}_k=\frac{1}{2m}\{p^i, g_{ik}\}
\end{equation}
The corresponding classical result leads directly, by inverting $g_{ik}$, to an expression for $p^i$ in terms of $\dot{x}_k$.
We can nevertheless invert equation 
(13). 
The anticommutator of $\dot{x}_k$ with $g^{kl}$ becomes
\begin{equation}
\begin{gathered}
\dot{x}_kg^{kl}+g^{kl}\dot{x}_k=\frac{1}{2m}[(p^ig_{ik}+g_{ik}p^i)g^{kl}+g^{kl}(p^ig_{ik}+g_{ik}p^i)]\\
=\frac{1}{2m}[p^i\delta_{il}+g_{ik}p^ig^{kl}+g^{kl}p^ig_{ik}+\delta_{il}p^i]\\
=\frac{1}{2m}[2p^l+p^ig_{ik}g^{kl}+[g_{ik}, p^i]g^{kl}+g^{kl}g_{ik}p^i+g^{kl}[p^i, g_{ik}]\\
=\frac{1}{2m}4p^l=\frac{2}{m}p^l
\end{gathered}
\end{equation}
so that
\begin{equation}
p^l=\frac{m}{2}\{\dot{x}_k,g^{kl}\}
\end{equation}
The cancellation in (14) that leads to this result follows from the fact that $[g_{ik},p^i]$ depends only on \textbf{x} and commutes with $g^{kl}$. This result implies that the Hamiltonian does not correspond to a simple bilinear form in $\dot{x}_i$, but a bilinear in anticommutators of the type
(15). As we shall see, this correspondence carries over to the analog of the geodesic type formula for $\ddot{x}_l$.
The classical form of eq. (15) (as in the classical theory) does not correspond to
the usual relation between momentum and velocity; moreover, by Eq. (14) the separate
components of $\dot{x}_k$  do not commute with each other, or, with $\{x_i\}$.
For example, it is straightforward to show that
\begin{equation}
[\dot{x}_k,x_l]=-\frac{i}{m}g_{lk}
\end{equation}
and
\begin{equation}
[\dot{x}_k,\dot{x}_l]=\frac{i}{2m^2}\{p^i,(g_{lm}\frac{\partial g_{ik}}{\partial x_m}-g_{km}\frac{\partial g_{il}}{\partial x_m})\}
\end{equation}
Thus, the  velocities  do not commute.
Let us define, in analogy to the classical case, a new set of operators (analogous to what were called $\{y\}$ in our discussion above of the classical case; here we 
 use the same notation)
\begin{equation}
\dot{y}^l\equiv \frac{1}{2}\{\dot{x}_k,g^{kl}\}
\end{equation}
so that, by (15),
\begin{equation}
p^l=m\dot{y}^l
\end{equation}
The $\{\dot{y}^l\}$ form a commutative set. Furthermore, it follows from (16) that
\begin{equation}
\begin{gathered}
$$
[\dot{y}^l,x_m]=\frac{1}{2}[\{\dot{x}_k,g^{kl}\}, x_m]=
g^{kl}[\dot{x}_k, x_m]=-\frac{i}{m}\delta^l_m
$$
\end{gathered}
\end{equation}
so that (consistently)
\begin{equation}
[p^l,x_m]=-i\delta_m^l
\end{equation}
We now turn to the second order equations for the dynamical variables. Since
from (19)
\begin{equation}
\begin{gathered}
\ddot{y}^l=-\frac{i}{m}[p^l,\hat{H}_G]=\\-\frac{i}{2m^2}p^i[p^l, g_{ij}(\textbf{x})]p^j=-\frac{1}{2m^2}p^i\frac{\partial g_{ij}}{\partial x_l}p^j
\end{gathered}
\end{equation}
Using (19) again, we arrive at the simple result
\begin{equation}
\ddot{y}^l=-\frac{1}{2}\dot{y}^i\frac{\partial g_{ij}}{\partial x_l}\dot{y}^j
\end{equation}
 closely related to
 the form obtained in the classical case \cite{horwitz} for the "geodesic" equation with reduced connection. In the classical case, this formula was used to compute geodesic deviation for the geometrical embedding of Hamiltonian motion (for Hamiltonian of the form $\frac{p^2}{2m}+V(\textbf{y})$) \cite{horwitz}, as discussed above, for which the corresponding metric was of the conformal form given in Eq. (3).\\
It follows from the Heisenberg equations applied directly to (18) that
\begin{equation}
\begin{gathered}
\ddot{y}^i=i[\hat{H}_G, \dot{y}^i]=\frac{i}{2}[\hat{H}_G, \dot{x}_kg^{ki}+g^{ki}\dot{x}_k]=\\\frac{1}{2}\ddot{x}_kg^{ki}+\frac{i}{2}\dot{x}_k[\hat{H}_G, g^{ki}]+\frac{i}{2}[\hat{H}_G, g^{ki}]\dot{x}_k+\frac{1}{2}g^{ki}\ddot{x}_k
\end{gathered}
\end{equation}
Computing the commutator ${[\hat{H}_G, g^{ki}]}$ in relation (24),
\begin{equation}
\begin{gathered}
{[\hat{H}_G, g^{ki}]}=\frac{1}{2m}[p^mg_{mn}p^n, g^{ki}]\\=\frac{1}{2m}\{[p^m, g^{ki}]g_{mn}p^n+p^mg_{mn}[p^n, g^{ki}]\}\\
=\frac{1}{2m}\{-i\frac{\partial g^{ki}}{\partial x_m}g_{mn}p^{n}+p^mg_{mn}(-i\frac{\partial g^{ki}}{\partial x_n})\}\\
{[\hat{H}_G, g^{ki}]}=-\frac{i}{2m}\{\frac{\partial g^{ki}}{\partial x_m}g_{mn}, p^n\}
\end{gathered}
\end{equation}
and therefore
\begin{equation}
\begin{gathered}
\ddot{y}^i=\frac{1}{2}\{\ddot{x}_k, g^{ki}\}+\frac{i}{2}\{\dot{x}_k, [\hat{H}_G, g^{ki}]\}\\
=\frac{1}{2}\{\ddot{x}_k, g^{ki}\}+\frac{1}{4m}\{\dot{x}_k, \{\frac{\partial g^{ki}}{\partial x_m}g_{mn}, p^n\}\}
\end{gathered}
\end{equation}
Furthermore, with relation (15), $p^n=\frac{m}{2}\{\dot{x}_k,g^{kn}\}$, relation (26) results in 
\begin{equation}
\begin{gathered}
\ddot{y}^i=\frac{1}{2}\{\ddot{x}_k, g^{ki}\}+\frac{1}{8}\{\dot{x}_k, \{\frac{\partial g^{ki}}{\partial x_m}g_{mn}, \{\dot{x}_a, g^{an}\}\}\}
\end{gathered}
\end{equation}
The relation between $\ddot{y}^i$ and $\ddot{x}_i$  therefore contains nonlinear velocity dependent inhomogeneous terms. It appears, however, that there should be  a strong relation between instability, sensitive to acceleration, in $\textbf{x}$ and $\textbf{y}$ variables.\\  
To find the second order equation for the operators $x_l$, we proceed directly  from (10), using (19), to write  the Hamiltonian as
\begin{equation}
	\hat{H}_G=\frac{m}{2}\dot{y}^ig_{ij}(\textbf{x})\dot{y}^j
\end{equation}
We may now compute
\begin{equation}
\ddot{x}_l=i[\hat{H}_G,\dot{x}_l]
\end{equation}
and use (17) and (18) to 
obtain the quantum mechanical form of the "geodesic" equation generated by the Hamiltonian $\hat{H}_G$:
\begin{equation}
\begin{gathered}
\ddot{x}_l=\frac{1}{16}(\{\{g^{nm},\dot{x}_m\},\frac{\partial g_{ln}}{\partial x_i}\},g_{ij}\{g^{jp},\dot{x}_p\}\\
-2\{g^{ip},\dot{x}_p\}g_{ln}\frac{\partial g_{ij}}{\partial x_n}\{g^{jq},\dot{x}_q\})
\end{gathered}
\end{equation}
In the classical limit, where all anticommutators become just (twice) simple products, a short computation  yields
\begin{equation}
\ddot{x}_l=-\Gamma^{pq}_l\dot{x}_p\dot{x}_q
\end{equation}
with
\begin{equation}
\Gamma^{pq}_l=\frac{1}{2}g_{ln}(\frac{\partial g^{nq}}{\partial x_p}+\frac{\partial g^{np}}{\partial x_q}-\frac{\partial g^{pq}}{\partial x_n})
\end{equation}
i.e., the classical geodesic formula generated by a classical Hamiltonian of the form (2) \cite{horwitz}. Therefore, (30) is a proper quantum generalization of the classical geodesic formula.\\
We finally express the quantum "geodesic" formula (30) explicitly in terms of the
canonical momenta using (18) and (19). 
We use the canonical commutation relations to write the result in terms of a bilinear in momentum ordered to bring momenta to the left and right. In this form we may describe the quantum state in terms of the variables $\{x_l\}$ canonically conjugate to the $\{p^j\}$ in the sense of (20-22), for which
\begin{equation}
p^j\rightarrow -i\frac{\partial}{\partial x_j}
\end{equation}
on $\psi(\textbf{x})$.\\
We then find that expression (30) for the quantum mechanical form of the "geodesic" equation $\ddot{x}_l$  could be written as
\begin{equation}
	\begin{gathered}
		\ddot{x}_l=\frac{1}{2m^2}p^i(\frac{\partial g_{li}}{\partial x_n}g_{nj}+\frac{\partial g_{lj}}{\partial x_n}g_{ni}-\frac{\partial g_{ij}}{\partial x_n}g_{ln})p^j\\
		+\frac{1}{4m^2}\frac{\partial}{\partial x_j}(\frac{\partial^2 g_{ln}}{\partial x_i\partial x_n}g_{ij})
	\end{gathered}
\end{equation}
The first term is closely related to the classical connection form; the second term, an essentially quantum effect.
\section{Local relation between the two coordinate bases}
We shall need some algebraic relations. 
Using  definition (19) and  relations (13) and (18), it follows that 
\begin{equation}
\begin{gathered}
$$
[\dot{y}^i,\dot{x}_l]=\frac{1}{2m}[\dot{y}^i,\{p^n, g_{nl}\}]\\=\frac{1}{2m}p^n[\dot{y}^i, g_{nl}]+\frac{1}{2m}[\dot{y}^i, g_{nl}]p^n
\\=-\frac{i}{2m^2}p^n\frac{\partial g_{nl}}{\partial x_i}-\frac{i}{2m^2}\frac{\partial g_{nl}}{\partial x_i}p^n\\
=-\frac{i}{2m}\{\dot{y}^n, \frac{\partial g_{nl}}{\partial x_i}\}
$$
\end{gathered}
\end{equation}
so that
\begin{equation}
[\dot{y}^i,\dot{x}_l]=-\frac{i}{2m}\{\dot{y}^n,\frac{\partial g_{nl}}{\partial x_i}\}
\end{equation}
Furthermore
\begin{equation}
\begin{gathered}
$$
[g_{ij},\dot{x}_l]=\frac{1}{2m}[g_{ij},\{p^n, g_{nl}\}]\\=\frac{1}{2m}[g_{ij}, p^n]g_{nl}+\frac{1}{2m}g_{nl}[g_{ij}, p^n]\\
=\frac{i}{m}g_{nl}\frac{\partial g_{ij}}{\partial x_n}  
$$
\end{gathered}
\end{equation}
so that
\begin{equation}
[g_{ij},\dot{x}_l]=\frac{i}{m}g_{nl}\frac{\partial g_{ij}}{\partial x_n}
\end{equation}
We also have that
\begin{equation}
\begin{gathered}
$$
[g_{ij}, \dot{y}^l]=\frac{1}{2}[g_{ij},\{\dot{x}_k, g^{kl}\}]\\
[g_{ij}, \dot{y}^l]=\frac{1}{2}[g_{ij}, \dot{x}_k]g^{kl}+\frac{1}{2}g^{kl}[g_{ij}, \dot{x}_k]
$$
\end{gathered}
\end{equation}
so that
\begin{equation}
\begin{gathered}
$$
[g_{ij}, \dot{y}^l]=g^{kl}[g_{ij}, \dot{x}_k]
$$
\end{gathered}
\end{equation}
Furthermore, the left term of relation (40) results in
\begin{equation}
\begin{gathered}
$$
[g_{ij}, \dot{y}^l]=[g_{ij}, i[\hat{H}_G, y^l]]\\
[g_{ij}, \dot{y}^l]=\frac{i}{2m}[g_{ij}, [p^mg_{mn}p^n, y^l]]\\
[g_{ij}, \dot{y}^l]=\frac{i}{2m}[p^m, y^l]g_{mn}[g_{ij}, p^n]+\frac{i}{2m}[g_{ij}, p^m]g_{mn}[p^n, y^l]
$$
\end{gathered}
\end{equation}
We assume that the operator $p^j$, on the Hilbert space representation $\chi(\textbf{y})\in\mathcal{H}_y$, constructed in accordance with the (Hermitian) operator form of (1), 
is represented by
\begin{equation}
p^j\rightarrow -i\frac{\partial}{\partial y^j}
\end{equation}
This assumption is analogous to the "dynamical equivalence" assumption of ref \cite{horwitz}, for which the classical momenta in the two descriptions are  taken to be identical for all time. The quantum mechanical momenta then generate translations in $\psi(\textbf{x})\in\mathcal{H}_x$ and $\chi(\textbf{y})$.\\
Therefore, applying relation (42) along with (33) to (41)  results in (as a formal definition)
\begin{equation}
\begin{gathered}
$$
[g_{ij}, \dot{y}^l]=\frac{i}{m}g_{ln}\frac{\partial g_{ij}}{\partial x_n}\equiv \frac{i}{m}\frac{\partial g_{ij}}{\partial y^l}
$$
\end{gathered}
\end{equation}
We may apply relation (42) to the right side of relation (40)
\begin{equation}
\begin{gathered}
g^{kl}[g_{ij}, \dot{x}_k]=\frac{1}{2m}g^{kl}[g_{ij}, p^n]g_{nk}+\frac{1}{2m}g^{kl}g_{nk}[g_{ij}, p^n]\\
=\frac{i}{m}\frac{\partial g_{ij}}{\partial y^l}
\end{gathered}
\end{equation} 
which is consistent with relation (43).  
As a result of relation (43) we  may think formally of a  transformation between the two  coordinate bases, $\{x_i\}$ and $\{y^j\}$, 
defined  locally by
\begin{equation}
\begin{gathered}
\frac{\partial g_{ij}}{\partial y^l}\equiv g_{ln}\frac{\partial g_{ij}}{\partial x_n}
\end{gathered}
\end{equation}
It should be emphasized that, in the 
classical case, Horwitz et al. \cite{horwitz} identified 
an effective embedding of the Hamiltonian dynamics into a non-Euclidean manifold, equipped with a connection form and a metric. 
Derivatives of functions on $\{y^i\}$ could then be related to derivatives of functions of $\{x_i\}$ \cite{yahalom}, but the global relation between these manifolds is not determined.
As a result of the differential relations found by \cite{yahalom}, the relation
\begin{equation}
\frac{\partial g_{ij}}{\partial x_m}=g^{mn}\frac{\partial g_{ij}}{\partial y^n}
\end{equation}
is, however, valid.
\section{Relation to Potential Model Hamiltonians}
As we have pointed out above, the 
acceleration satisfies the "geodesic"
type equation (23). The truncated connection form in this equation is 
similar in form to that
derived for the classical case and used in a large number of applications \cite{yossi} to generate, by geodesic deviation, a criterion \cite{horwitz} for stability of a Hamiltonian of the potential model form (1)
\begin{equation}
H=\delta_{ij}\frac{p^ip^j}{2m}+V(\textbf{y})
\end{equation}
where we have assigned the variable \textbf{y} to correspond to the configuration space of the potential model.\\
For a Hamiltonian of form (1), we wish to make a correspondence, as in the classical case, with a geometric Hamiltonian of the form (10). Let us suppose that $g_{ij}(\textbf{x})$ has the form (3) for this case as well. Then,
\begin{equation}
\hat{H}_G=\frac{1}{2m}p^i\Phi(\textbf{x})p^j\delta_{ij}
\end{equation}
In the semiclassical limit, we choose states in $\mathcal{H}_\textbf{x}$ and $\mathcal{H}_\textbf{y}$ that are localized in $p^i$, and, for consistency in the application of the Ehrenfest approximation \cite{ehrenfest}, fairly well localized in $\textbf{x}$ and $\textbf{y}$ as well (within the uncertainty bonds). Then, for such a wave function
\begin{equation}
<\hat{H}_G>_{\chi_\textbf{p}}\approxeq \frac{1}{2m}\textbf{p}^2\Phi(\textbf{x})
\end{equation}
and
\begin{equation}
<\hat{H}>_{\psi_\textbf{p}}\approxeq\frac{1}{2m}\textbf{p}^2+V(\textbf{y})
\end{equation}
Assigning a common value E to $<\hat{H}_G>_{\chi_\textbf{p}}$ and $<\hat{H}>_{\psi_\textbf{p}}$, we see that the choice (as in (4))
\begin{equation}
\Phi(\textbf{x})=\frac{E}{E-V(\textbf{y})}
\end{equation} 
is effective in this approximation as well. The relation between the functions on the left and right hand sides established by (45) and ref.\cite{yahalom} is valid in this contest as well.
We now return to (23), which can be written, with the help of (45), as
\begin{equation}
\ddot{y}^l=-\frac{1}{2}\dot{y}^ig^{ln}\frac{\partial g_{ij}}{\partial y^n}\dot{y}^j
\end{equation} 
Using the relation (51), this becomes
\begin{equation}
\ddot{y}^l=-\frac{1}{2}\dot{y}^i\frac{1}{E-V(\textbf{y})}\frac{\partial V(\textbf{y})}{\partial y^l}\dot{y}^j\delta_{ij}
\end{equation} 
so that (we do not distinguish upper and lower indices of $y^l$ here)
\begin{equation}
\ddot{y}^l=-\frac{1}{2m^2}p^i\frac{1}{E-V(\textbf{y})}\frac{\partial V(\textbf{y})}{\partial y^l}p^j\delta_{ij}
\end{equation}
The expectation values of $\ddot{y}^l$ in the state $\chi_\textbf{p}$ is the
\begin{equation}
<\ddot{y}^l>_{\chi_\textbf{p}}\approxeq-<\frac{\partial V(\textbf{y})}{\partial y^l}>_{\chi_\textbf{p}}
\end{equation}
recovering, in this semiclassical limit, the quantum evolution of the particle in the Ehrenfest approximation. 
Thus, as pointed out above, a calculation of "geodesic deviation" based on the quantum formula (23) could provide a new criterion for quantum chaos, consistent with the Bohigas conjecture \cite{bohigas} due to the similar  structure of the classical 
and quantum criteria (see below).
\section{Stability Relations as a Criteria for Unstable Behavior}
Following the classical notion of inducing an infinitesimal translation of a trajectory around a given initial point along the trajectory we refer to a corresponding quantum mechanical notion of "geodesic deviation" based on the quantum formula (23).\\
For "geodesic deviation" we then induce a translation as follows (where we define \textbf{$\xi$} as a common number);
\begin{equation}
\psi_t(\textbf{x})\rightarrow\psi_t(\textbf{x}+\textbf{$\xi$})
\end{equation}
that is
\begin{equation}
\psi_t(\textbf{x}+\textbf{$\xi$})=e^{\frac{ip^l\xi^l}{\hbar}}\psi_t(\textbf{x})
\end{equation}
Computing $\delta(\psi_t,\ddot{y}^l\psi_t)(t)$, assuming now the physical system is supposed almost classical, results in
\begin{equation}
-\frac{1}{2}<\psi_t|\dot{y}^i(\frac{\partial}{\partial x_a}(\frac{\partial g_{ij}}{\partial x_l}))\dot{y}^j|\psi_t>\xi^a\stackrel{def}{=}\ddot{\xi^l}(t)
\end{equation}
where the left side of expression (58) is \textit{defined} as the second derivative with respect to
the common number $\xi^l$, the distance between the two trajectories as a function of time, 
 which is expected to coincide
in the Ehrenfest approximation \cite{ehrenfest} with 
Horwitz et al. study \cite{horwitz}.
Substituting relation (45) for a local transformation between the two  coordinate basis $\{x_i\}$ and $\{y^j\}$ results in
\begin{equation}
\begin{gathered}
-\frac{1}{2}<\psi_t|\dot{y}^i(g^{am}\frac{\partial}{\partial y^m}(g^{ln}\frac{\partial g_{ij}}{\partial y^n}))\dot{y}^j|\psi_t>\xi^a\stackrel{def}{=}\ddot{\xi^l}(t)
\end{gathered}
\end{equation}
Thus one may write expression (59) as the sum of two terms
\begin{equation}
\begin{gathered}
-\frac{1}{2}<\psi_t|\dot{y}^i(g^{am}\frac{\partial g^{ln}}{\partial y^m}\frac{\partial g_{ij}}{\partial y^n})\dot{y}^j|\psi_t>\xi^a\\-\frac{1}{2}<\psi_t|\dot{y}^i(g^{am}g^{ln}\frac{\partial^2 g_{ij}}{\partial y^m\partial y^n})\dot{y}^j|\psi_t>\xi^a
\stackrel{def}{=}\ddot{\xi^l}(t)
\end{gathered}
\end{equation}
We then define
\begin{equation}
\hat{\ddot{{\xi}}}^{al}\stackrel{def}{=}-\frac{1}{2}\dot{y}^ig^{am}(\frac{\partial g^{ln}}{\partial y^m}\frac{\partial g_{ij}}{\partial y^n}+g^{ln}\frac{\partial^2g_{ij}}{\partial y^m\partial y^n})\dot{y}^j
\end{equation}
which we call  the \textbf{operator geodesic deviation}.
The expectation values associated with this operator correspond  in the Ehrenfest approximation \cite{ehrenfest} to a measure of deviation between 
the expectation values of two neighboring evolutions of $y^l$
which in the classical analogy are  characteristic of the geodesic deviation between two near by trajectories in space. In the Ehrenfest approximation they are expected to determine the dynamic flow of the  position variable's expectation values.
 Based on the  sensitivity to the local instability criterion
  it is expected   that  in  Ehrenfest approximation 
 one can characterize the "local" stability properties of the dynamic flow of the coordinate expectation variables.
We  use these matrix coefficients's eigenvalues to determine "local" instability. 
 If one of them is negative, it is sufficient to imply "local" instability.
One can map out the regions of instability over the physically admissible region using these formulas.
We shall also follow the orbits to see their behavior, as exhibited by the expectation values. We
have found, so far, a remarkable correlation between the simulated orbits and the predictions of "local"
instability in this way. Note that these curves of the expectation values of $y^i$
do not necessarily correspond to classical particle trajectories in the sense of Ehrenfest. As pointed out above, the
Ehrenfest theorem fails after some time \cite{zaslavsky} . We see, however, that the expectation values contain
important diagnostic behavior \cite{ballentine}, and could well be incorporated into a new definition of "quantum
chaos", corresponding to deviation under small perturbation (change of initial conditions).
Although the Ehrenfest correspondence fails rapidly in case of chaotic behavior, 
from the behavior of the solutions it appears that this criterion may nevertheless provide a good definition for
quantum chaotic behavior. The correspondence between the classical and quantum definitions
provides, furthermore, support for the Bohigas conjecture \cite{bohigas}, i.e., that a classical Hamiltonian
generating chaotic dynamics goes over to a quantum theory exhibiting characteristics of chaotic
quantum behavior. Our simulations indicate that this will be true, for the examples we consider below.\\
We will show below through simulation by numerical analysis that this formula works well beyond the Ehrenfest approximation.
\section{Quantum dynamics stability properties vs. classical dynamics stability properties}
In this section we wish to define 
a manifold  which  we suggest as  corresponding to the \textbf{classical} \textit{Gutzwiller manifold} presented by \cite{horwitz}. We then work through an  analytical analysis to  investigate the stability properties of the trajectories covering this manifold and we
derive local stability criterions which we assume to 
be related to 
the stability behaviors of the corresponding quantum mechanical dynamics. 
We will  show some computational 
 examples in which applying these stability criteria predicts correctly the stability properties of the quantum  dynamics. 
This might suggest that 
the trajectories have
the same stability properties as in the underlying quantum case and through the local criterion it seems that one may predict 
chaotic quantum mechanical behavior as well.\\ 
To understand the geometric contest of our formality let us define a manifold  as  follows \cite{yossi}:
\begin{itemize}
	\item Let $\mathcal{H}$ be a Hilbert space corresponding to a given quantum mechanical system and let $H_G$ be the self-adjoint Geometric Hamiltonian generating the evolution of the system.
	\item Let $\varphi(t)=U(t)\varphi=e^{-iH_Gt}\varphi$
	be the state of the system at time t corresponding to an initial \textbf{localized} state $\varphi(0)=\varphi\in\mathcal{H}$ such that (as long as) $\varphi(t)$ is \textbf{localized} too, i.e.
	the width of a well-localized state is extremely narrow compared to the characteristic variation in the matrix function $g_{ij}(\textbf{x})$ which is the  relevant system dimension.
	\item Define the \textit{trajectory} $\Phi_{\varphi}$ corresponding to an initial \textit{localized} state $\varphi\in\mathcal{H}$ to be\\
	$\Phi_{\varphi}:=\{U(t)\varphi\}_{t\in R^+}=\{\varphi(t)\}_{t\in R^+}$\\
	i.e.,$\Phi_{\varphi}$ is the set of \textbf{localized} 
	states reached in the course of the evolution of the system from an initial \textit{localized} state $\varphi$. 
	\item Next, let us denote by\\ $M(\Phi_{\varphi})=\{(\varphi,M\varphi)|\varphi\in\Phi_{\varphi}\}$\\
	the collection of all expectation values of $M$ for \textit{localized} states in $\Phi_{\varphi}$ (up to a multiplicative normalization constant).
	\item Next, let us denote by\\ $\{M(\Phi_{\varphi})\}_{\Phi_{\varphi}\in\mathcal{H}}=\{M(\Phi_{\varphi})|\Phi_{\varphi}\in\mathcal{H}\}$\\
	the collection of all expectation values of $M$ for all possible trajectories $\Phi_{\varphi}$ in $\mathcal{H}$, i.e. $\{M(\Phi_{\varphi})\}_{\Phi_{\varphi}\in\mathcal{H}}\subseteq R^n$ where $R^n$ is the n-dimensional Euclidean space. We call $\{M(\Phi_{\varphi})\}_{\Phi_{\varphi}\in\mathcal{H}}$ the \textit{expectation values manifold}.
\end{itemize}
 The classical limit 
 of the corresponding collection of all possible trajectories evolved in this way in the course of a long period of time appear to be valid beyond localization \cite{ballentine}.\\ 
Consider the coordinate bases $\{x_i\}$ and $\{y^j\}$. 
Given relation (45), we may think of the transformation matrix $A=\{a^j_i\}$
between the two sets of bases vectors \cite{guera}
\begin{equation}
\hat{e}_i\equiv\{\frac{\partial}{\partial x_i}\}\rightarrow
 \hat{e}^j\equiv\{a^j_i\frac{\partial}{\partial x_i}\}\equiv\{g_{ji}\frac{\partial}{\partial x_i}\}
\end{equation} 
thus the set $\{\hat{e}^j\}$ forms another  set of basis vectors associated with the coordinate basis defined previously $\{y^j\}$. Then we have a new set of basis vectors locally for each tangent space on the manifold.
We would like our new frame to be
orthonormal at all points. That is,
\begin{equation}
g(\hat{e}_i, \hat{e}_j)=g(a_i^n\hat{e}_n, a_j^k\hat{e}_k)=a_i^na_j^kg_{nk}=\delta_{ij}
\end{equation}
This equation can be rearranged
\begin{equation}
a_i^ng_{nk}(a_k^j)^T=\delta_{ij}\rightleftarrows AGA^T=I
\end{equation}
We get the matrix (we choose
A to be symmetric)
solution $A=G^{-\frac{1}{2}}$.
If the matrix
component, $g_{ik}({\textbf{x}})$, is considered as a  slowly varying  function on the coordinates basis in the position representation then we can expand this solution in a power series around $G=I$
\begin{equation}
\begin{gathered}
A=G^{-\frac{1}{2}}=I+\sum_{n=1}^{\infty}\frac{(-1)^n1\cdot3...(2n-1)}{n!\cdot2^{n}}\tilde{G}^n\\
where\quad \tilde{G}\stackrel{def}{=}G-I\\
\end{gathered}
\end{equation}
In terms of components 
the first few terms are;
\begin{equation}
a^j_i=\delta_{ij}-\frac{1}{2}\tilde{g}_{ij}+\frac{3}{8}\sum_{k=1}^{N}\tilde{g}_{ik}\tilde{g}_{kj}-...
\end{equation}
Substituting the zero order  of the matrix expansion  applied to  $g^{am}$ in relation (60) where  the localization construction is required to be valid  results in 
\begin{equation}
\begin{gathered}
{\ddot{{\xi}}}^{l}\approx-\frac{1}{2}\dot{y}^i\delta^{am}(\frac{\partial g^{ln}}{\partial y^m}\frac{\partial g_{ij}}{\partial y^n}+g^{ln}\frac{\partial^2g_{ij}}{\partial y^m\partial y^n})\dot{y}^j\xi^a\\=-\frac{1}{2}\dot{y}^i(\frac{\partial g^{ln}}{\partial y^a}\frac{\partial g_{ij}}{\partial y^n}+g^{ln}\frac{\partial^2g_{ij}}{\partial y^a\partial y^n})\dot{y}^j\xi^a
\end{gathered}
\end{equation}
This zero order expansion  is \textit{precisely}  the form obtained in the    classical case  for the  second order geodesic deviation equations (considering the second term only of a zero order) (see eq. (22) in \cite{horwitz}). It is implied that in the quantum case the orbit deviation equations become oscillatory.\\
Thus one may conclude that   in the classical limit  the  first  order of  the matrix expansion (66)  applied to $g^{am}$ in relation (60) may be considered as a quantum mechanical effect
\begin{equation}
\begin{gathered}
-\frac{1}{2}\dot{y}^i\tilde{g}^{am}(\frac{\partial g^{ln}}{\partial y^m}\frac{\partial g_{ij}}{\partial y^n}+g^{ln}\frac{\partial^2g_{ij}}{\partial y^m\partial y^n})\dot{y}^j\xi^a
\end{gathered}
\end{equation}
which could be viewed as the low order expansion of expression (60) around the classical relation of a  "geodesic" deviation equation \cite{horwitz}.\\
Then the low orders expansion of the quantum mechanical "geodesic deviation" in the classical limit results in
\begin{equation}
\begin{gathered}
\ddot{\xi^l}(t)\approx-\frac{1}{2}\dot{y}^i(\frac{\partial g^{ln}}{\partial y^a}\frac{\partial g_{ij}}{\partial y^n}+g^{ln}\frac{\partial^2g_{ij}}{\partial y^a\partial y^n})\dot{y}^j\xi^a\\
-\frac{1}{2}\dot{y}^i\tilde{g}^{am}(\frac{\partial g^{ln}}{\partial y^m}\frac{\partial g_{ij}}{\partial y^n}+g^{ln}\frac{\partial^2g_{ij}}{\partial y^m\partial y^n})\dot{y}^j\xi^a
\end{gathered}
\end{equation}
In the classical limit
one may refer to expression (69)  as 
a measure of deviation between the two dynamics of $y^l$'s expectation values 
which becomes the corresponding quantum  second order "geodesic deviation" equations.
On this view, the quantum dynamics may be identified as
chaotic provided the flow of the quantum expectation values approximately follow a chaotic classical trajectory.\\ 
Moreover, one may claim that 
in the classical limit expression (69) is
expected to 
admit 
the emerging of a corresponding quantum "local" criterion for
stability properties of the dynamical flow of the  position variable's expectation values 
in correspondence with
the  sensitivity to the local 
stability criterion for unstable behavior derived by \cite{horwitz} in the classical case.\\
Then one may write  in the classical limit a corresponding quantum mechanical "local" criterion which reflects the stability of the trajectories associated with the position expectation values (average particle's position). This view may be expressed as
\begin{equation}
\ddot{\xi}(t)=-(\mathcal{V})P\xi
\end{equation}
where
\begin{equation}
\begin{gathered}
\mathcal{V}^l_{ija}\stackrel{def}{=}\frac{1}{2}(\frac{\partial g^{ln}}{\partial y^a}\frac{\partial g_{ij}}{\partial y^n}+g^{ln}\frac{\partial^2g_{ij}}{\partial y^a\partial y^n})\\
+\frac{1}{2}\tilde{g}^{am}(\frac{\partial g^{ln}}{\partial y^m}\frac{\partial g_{ij}}{\partial y^n}+g^{ln}\frac{\partial^2g_{ij}}{\partial y^m\partial y^n})\\
\end{gathered}
\end{equation}
and $P^{ij}=\delta^{ij}-\frac{<\dot{y}^i><\dot{y}^j>}{<\dot{\textbf{y}}>^2}$, 
defining a projection into a direction orthogonal
to the average velocity, $<\dot{\textbf{y}}>$, i.e.  the component orthogonal to the flow of the position expectation values.\\
"Local" instability should occur if at least one of the eigenvalues of the matrix $\mathcal{V}^l_{ija}$
is \textit{negative}.\\
We define for brevity
\begin{equation}
\begin{gathered}
{\mathcal{C}}_{ija}^{l}\stackrel{def}{=}\frac{1}{2}(\frac{\partial g^{ln}}{\partial y^a}\frac{\partial g_{ij}}{\partial y^n}+g^{ln}\frac{\partial^2g_{ij}}{\partial y^a\partial y^n})\\
 {\mathcal{Q}_{ija}}^{l}\stackrel{def}{=}\frac{1}{2}\tilde{g}^{am}
(\frac{\partial g^{ln}}{\partial y^m}\frac{\partial g_{ij}}{\partial y^n}+g^{ln}\frac{\partial^2g_{ij}}{\partial y^m\partial y^n})
\end{gathered}
\end{equation}
Since the matrix $\mathcal{V}$ embeds in its structure a sum of two terms, i.e. expression (71), where the first one as mentioned previously carries the precise structure as the matrix involved in the classical case \cite{horwitz}, while the second term is a quantum effect, then it is implied by equation (70) that the position expectation trajectories  may demonstrate "local" instabilities while the classical corresponding trajectory defined by Hamilton's equations of motion is stable and conversely.\\
Those cases may be expressed by the following  inequalities  for "local" instabilities. Given $\{\lambda_1, \lambda_2, \lambda_3, ..., \lambda_n\}$ are the matrix eigenvalues of $\mathcal{C}_{ija}^{l}$ and $\{\alpha_1, \alpha_2, \alpha_3, ..., \alpha_n\}$ are the matrix eigenvalues of $\mathcal{Q}_{ija}^{l}$, then 
\begin{equation}
\begin{gathered}
if\quad\exists(\lambda_l, \alpha_l)\quad such\quad that\\
0\leq \alpha_l<(-\lambda_l)\\
\textit{quantum instability is in correspondence with}\\ \textit{classical instability}
\end{gathered}
\end{equation}
But	
\begin{equation}
\begin{gathered}
if\quad\exists(\lambda_l, \alpha_l)\quad such\quad that\\ 
0\leq\lambda_l<(-\alpha_l)\\
 \textit{classical stability vs. quantum instability may emerge}
 \end{gathered}
 \end{equation}
while
\begin{equation}
\begin{gathered}
if\quad\exists(\lambda_l, \alpha_l)\quad such\quad that\\
0<(-\lambda_l)\leq\alpha_l\\
 \textit{quantum stability vs. classical instability may emerge}
\end{gathered}
\end{equation} 
As Zaslavsky \cite{zaslavsky} has pointed out, however, the Ehrenfest correspondence fails rapidly in case of chaotic behavior.
Nevertheless, 
Ballentine, Yang, and Zibin \cite{ballentine} compared quantum
expectation values and classical ensemble averages for the low-order moments
for initially localized states.
This correspondence suggests that the statistical properties of Liouville
mechanics continue to last even after Ehrenfest's approximation fails.
The correspondence
with Liouville mechanics, given fixed system size, was shown to be 
accurate for a longer time period under conditions of classical chaos.\\
In the Ehrenfest approximation and beyond, i.e. even after Ehrenfest correspondence fails in case of chaotic behavior, the collection $\textbf{y}(\Phi_{\varphi})$ 
will satisfy dynamical equations closely related to those for which the classical 
 ensemble averages describe the possible configurations for a classical system in phase space. 
\section{Simulations}
In this section we study the stability characters of the quantum dynamics through numerical simulations.
We present here three approaches. One which uses a direct time evolution of the wave packet simulated by the Schroedinger time dependent equation. In each time step, the position  expectation values are computed, directly, by the wave function (using a narrow coherent state wave packet).
Moreover, each simulation is repeated by a small change in the initial condition (small change in the initial average momentum of the wave packet) and the two trajectories are computed. The following diagrams (fig.1 and fig.2) show the orbits of the particles in the presence of the regions of instability described above. One sees that the instability of the orbits, demonstrated by crossing trajectories and complexity, reflects the divergence of nearby trajectories which can lead to chaotic behavior.
\\In the second approach  
we are searching for "local" instabilities as they are defined through the \textbf{operator geodesic deviation}, $\hat{\ddot{\xi^{al}}}$ (eq. (61)), i.e. the expectation values associated with this operator are characterizing in the Ehrenfest approximation a measure of deviation between two dynamic flows of $\ddot{y}^l$'s expectation values  which in the classical analogy are a characteristic of a geodesic deviation between two nearby trajectories in space and "local" instability will be defined by computing the matrix's eigenvalues where if one of them is negative, it is sufficient to determine
"local" instability.
In the following, we display a set of graphs (fig.3) showing the unstable regions. It is expected that the
trajectories, i.e. $<\textbf{y}>$, of the particles will be deflected strongly when they pass through these unstable
regions, causing instability of the motion \cite{yossi}.
We use as initial wave function the coherent state wave packet :
\begin{equation}
\psi(x,y,0)=\frac{1}{\pi}e^{-\frac{1}{2}(x-x_0)^2+ip^0_xx}e^{-\frac{1}{2}(y-y_0)^2+ip^0_yy}
\end{equation}
where
\begin{equation}
\begin{gathered}
<x>=x_0\quad <p_x>=p^0_x\\
<y>=y_0\quad <p_y>=p^0_y
\end{gathered}
\end{equation}
We examine  simulations of a set of five quantum 2D  wells while increasing the energy hypersurface, approaching the separatrix of the centered well. 
As can be easily observed, the centered well becomes an unstable dynamical region as the threshold energy is approached. This phenomena takes place due to the quantum effect terms which seem 
to "sense" the unstable regions outside the accessible classical region. The same behavior is observed also for the rest of the wells. 
 Moreover, one may think of the quantum mechanical terms as  acting as a "driving force"  (the system, as will be shown shortly, could be thought of  as a driven quantum 2D five well anharmonic oscillator)  which is quite significant due to the energy interval range of influence as it is manifested in the following simulations (fig.(3)).
It is suggested that  the "fluctuating" term  of the "local curvature"   of the energy hypersurface might be quite significant near the separatrix of the wells. 
This would turn the undriven system  to be effectivlly  a driven dynamics with the same stability properties and dynamic behavior. 
 As Lin and Ballentine  pointed \cite{lin}, the tunneling rate of an electron through a semiconductor double quantum well structure can be enhanced  by means of a dc bias or ac field (i.e. a driven anharmonic oscillator). Our work implies that it is reasonable to assume that   also  by approaching to the separatrix of a well tunneling might be enhanced.
This phenomena we are presenting here has never been previously observed.
Further research is required to confirm this idea. 
 In the third approach we continue to examine the instability inequalities relations (eq.73-75). We follow the work of Feit and Fleck \cite{feit} which simulated wave packet dynamics and chaos in the Henon-Heiles system. The evolution of wave packets under the influence of a Henon-Heiles potential was investigated in \cite{feit} by direct numerical solution of the time-dependent Schroedinger
equation, i.e. coherent state Gaussians with a variety of mean positions and
momenta were selected as initial wave functions. Four cases were simulated by \cite{feit};  two of the wave packets can be reasonably judged to exhibit regular or nonchaotic behavior and the remaining two chaotic behavior. 
All of the four cases exhibit corresponding \textbf{classical} motions which are \textbf{regular}.\\
Following the instability inequalities relations  we are able to derive \textit{analytically} (eqs.(73)-(75)) those simulated results where the conformal structure (eq.3-4) has been applied.  Those analytic results are summarized in table 1 below.
We wish to emphasize that case (d) reflects our definition of "local" instability as an appropriate definition for a quantum mechanical analogy of the classical notion of "local" instability since it corresponds to the Feit and Fleck \cite{feit} findings for case (d) where the behavior at early times appears regular.\\
\begin{tabular}{|l|c|r|}
	\hline
	\multicolumn{3}{|c|}{Table 1: Instability Inequality Criterions}\\
	\hline
	Case & Instability Inequality Relations & Behavior\\
	\hline case (a) & $0\leq (-\alpha_l)\leq\lambda_l$ & regular \\
	case (b) &$0\leq (-\alpha_l)\leq\lambda_l$ & regular \\
	case (c) & $0\leq\lambda_l<(-\alpha_l)$ & chaotic \\
	case (d) &$0\leq (-\alpha_l)\leq\lambda_l$ & regular \\
	\hline
\end{tabular}
\section{Conclusions}
We have introduced a new generalized analytic approach and on this basis we have derived a new diagnostic tool for identifying a quantum chaotic behavior. This might
serves as a new definition for quantum Hamiltonian chaos. We have shown results of simulation of the quantum motion in two 
dimensions,  showing in certain cases that the transition
 from predominantly regular to predominantly irregular trajectories (of expectation values) takes place over a narrow energy range about some energy Ec. We have shown an independent stability behavior of  the quantum dynamics from the classical correspondence. We were able to predict local stability properties depending on energy, but even more surprising, on the initial conditions of the wave packets and identifying zones of local stability and instability (on the level of the space expectation values). Moreover, this last demonstrates a  possible KAM-like transition,
which by itself could serve as a new criterion and definition for quantum chaos very much like in the classical case. 

\onecolumn
\begin{figure}[h]
	\centering
	\mbox{\subfigure[]{\epsfig{figure=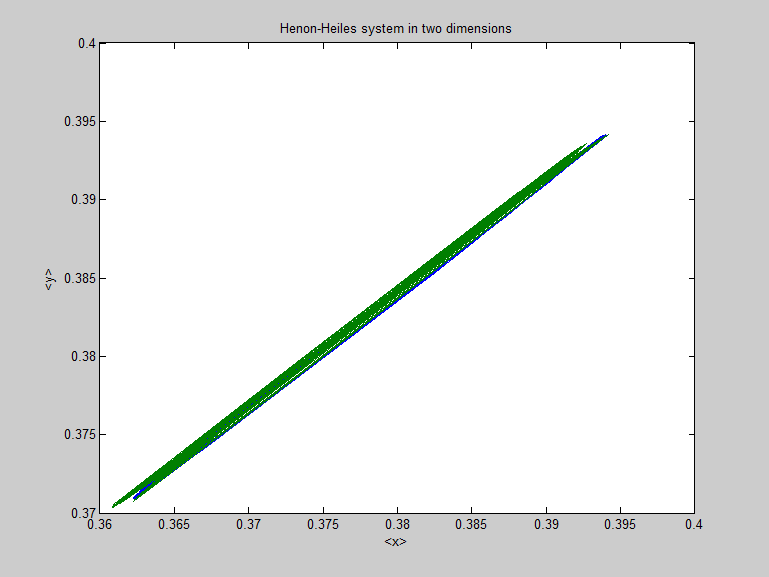,width=2.5in}}\quad
		\subfigure[]{\epsfig{figure=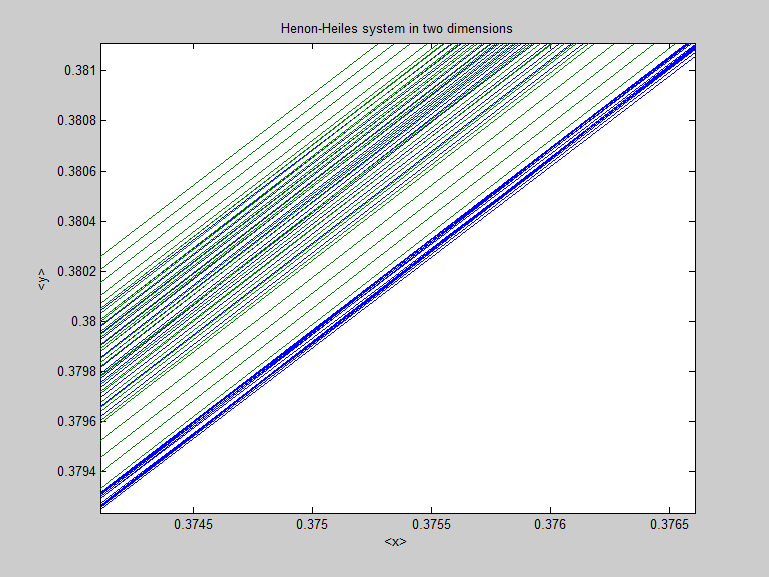,width=2.5in}}}
	\caption{(a) Henon-Heilles quantum dynamical simulation corresponding to a low classical energy, in a range known to be a non chaotic for classical dynamic. The two trajectories (blue and green) are closely together, reflecting no divergence of nearby trajectories, which  indicates, on the basis of our theory, a stable dynamics. (b) Zoomed picture of fig.1 above, presenting parallel trajectories, no complexity and nearby trajectories, indicating stability of the wave packet dynamics.}
	\label{fig1}
\end{figure}
\begin{figure}[h]
	\centering
	\includegraphics[width=2.5in]{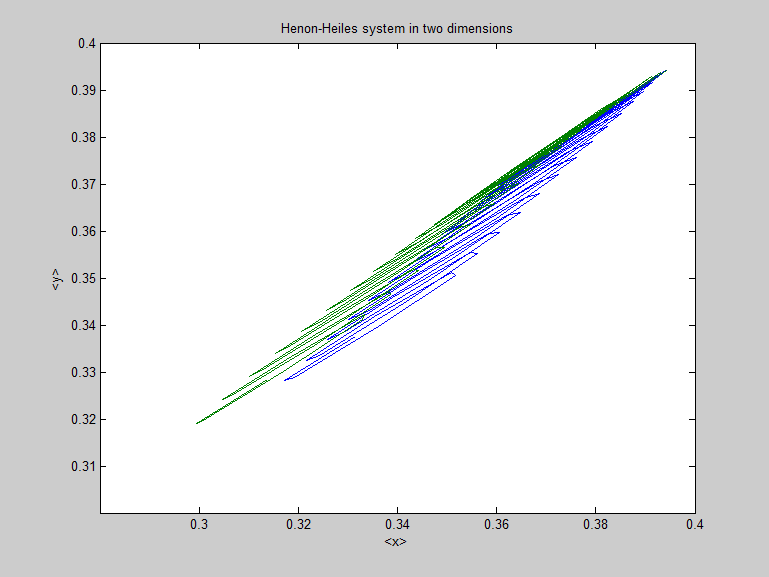}
	\caption{Henon-Heilles quantum dynamical simulation corresponding to a higher classical energy, in a range known to be a chaotic classical dynamics. The two trajectories starting from the same initial position's expectation value, separate along the time evolution of the dynamics, presenting broken trajectories (the trajectories are crossing themselves) and complexity indicating a chaotic dynamical behavior}
	\label{fig3}
\end{figure}
\clearpage
\begin{figure}[h!]
	\centering
	\mbox{\subfigure[]{\epsfig{figure=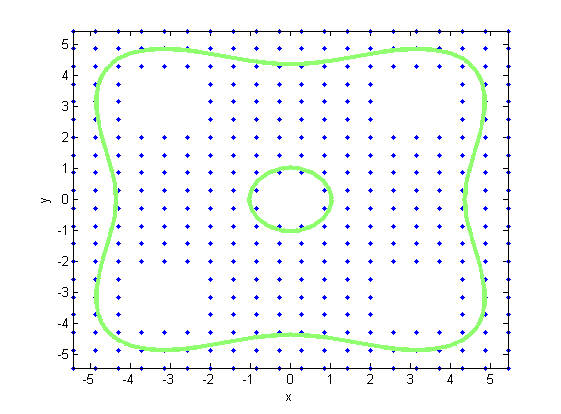,width=2.5in}}\quad
		\subfigure[]{\epsfig{figure=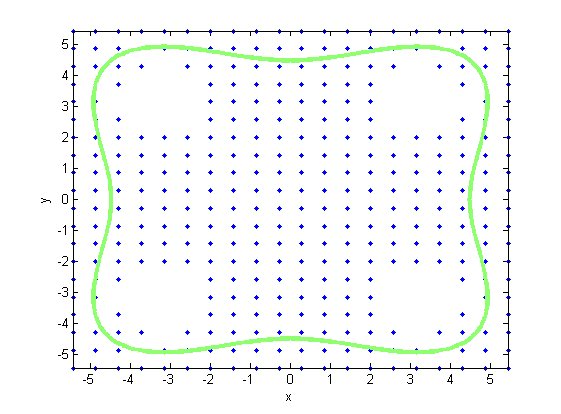,width=2.5in}}}
	\caption{(a) Five 2D  well  dynamical stability analysis for E=-36.6709. Stable dynamical regions are presented through 5 wells, separated by highly chaotic areas.
		Green lines indicating the classical hypersurface of classical energy as was discussed previously.
	(b) Five 2D  well  dynamical stability  analysis for the separatrix of the centered well.}
	\label{fig4}
\end{figure}
\begin{figure}[h]
	\centering
	\includegraphics[width=2.5in]{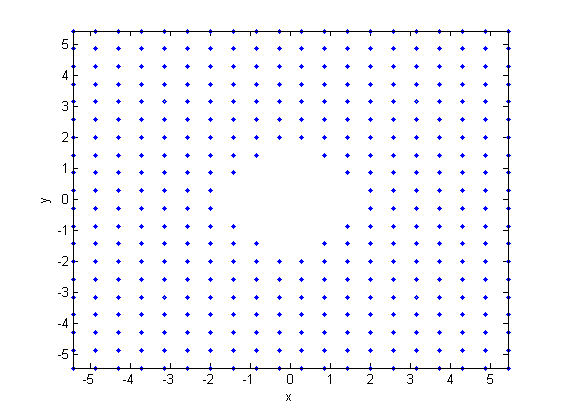}
	\caption{Five 2D  well  dynamical stability  analysis for the separatrix of the  off-centered wells.}
	\label{fig3}
\end{figure}


 


\end{document}